\documentclass{ws-ijmpa}
\usepackage{amssymb,graphicx,amsmath,graphics,subfigure,xfrac,url,xcolor}

\newcommand\inc{\mathrm{inc}}
\newcommand\reff{\mathrm{ref}}
\newcommand\trans{\mathrm{trans}}

\begin{document}

\title{The superradiance phenomenon in spin-one particles}

\author{Sebasti\'an Valladares}
\address{School of Physical Sciences and Nanotechnology, Yachay TECH University, 100119-Urcuqu\'i, Ecuador.\\
sebastian.valladares@yachaytech.edu.ec}

\author{Clara Rojas}
\address{School of Physical Sciences and Nanotechnology, Yachay TECH University, 100119-Urcuqu\'i, Ecuador.\\
crojas@yachaytech.edu.ec}

\maketitle

\pub{Received (\today)}{Revised (Day Month Year)}

\begin{abstract}

In this article, we solve the Duffin--Kemmer--Petiau (DKP) equation in the presence of hyperbolic tangent potential for spin-one particles. By partitioning the spin-one spinor, we show that the DKP equation is equivalent to the Klein--Gordon equation formalism. The scattering solutions are derived in terms of hypergeometric functions. The reflection $R$ and transmission $T$ coefficients are calculated in terms of the Gamma functions. The results show the presence of the superradiance phenomenon when $R$ for a specific region in the potential becomes greater than one.

\end{abstract}

\ccode{PACS Nos.: 02.30.Gp, 03.65.Pm, 03.65.Nk}

%%%%%%%%%%%%%%%%
\section{Introduction}	
%%%%%%%%%%%%%%%%

The study of relativistic wave equations has been of great significance. It allows us to understand several physical phenomena of Relativistic Quantum Mechanics, such as bound states, transmission resonances, and superradiance. The Dirac equation is not the only first-order relativistic equation in the literature another first-order relativistic equation can be found, namely, the Duffin--Kemmer--Petiau (DKP) equation\cite{falek2010duffin}. These two equations are closely related; in fact, they shared the same structure but replace the gamma matrices with beta matrices which follow a more complex algebra, namely the DKP algebra\cite{hamzavi2013approximate}. In contrast to the Dirac equation, the DKP equation describes vector (spin-one), and scalar (spin-zero) particles\cite{sogut:2010}. Remarkably, when the DKP equation is under the influence of a one-dimensional potential, the formalism for the spin-one is similar to the spin-zero formalism\cite{cardoso2010effects}.

With possible applications in several areas of study (as in nuclear and particle physics, cosmology, meson spectroscopy, and even nuclear-hadron interactions), the analytical solution of the DKP  equation for different potential wells and barriers has generated considerable interest in recent years\cite{chetouani:2004,merad:2007,de2010bound,sogut:2010,hassanabadi:2012,hassanabadi:2013a,hassanabadi:2013b,bahar:2103,darroodi:2015,darroodi2017exact,langueur2019dkp,langueur2021dkp,hamil:2021}. In particular, the DKP equation provides a broad background when studying spin-one particle interactions\cite{bahar2013aim}. In contrast to conventional descriptions based on the second-order Klein-Gordon and Proca equations, the DKP equation uses a first-order relativistic equation to study spin--zero and spin--one particles in a unified fashion\cite{chargui2013confinement}. As far as interactions are concerned, the DKP theory offers the advantage of being richer and allows for a wide variety of couplings that Klein–Gordon and Proca's theories cannot.

The superradiance phenomenon also has been widely studied\cite{chetouani:2004,de2010bound,manogue1988klein,rojas:2015,boutabia2005solution} and is a well-known phenomenon that occurs for steps barriers in Relativistic Quantum Mechanics\cite{de2010bound}. When Superradiance occurs, the reflection coefficient is greater than one ($R>1$), and hence the transmission coefficient is lower than zero ($T<0$). Nonetheless, the sum of both still accomplishes the unitary relation $T+R=1$.  This phenomenon has been studied in the Klein--Gordon equation\cite{rojas:2015}, the Dirac equation\cite{manogue1988klein}, and the DKP equation\cite{chetouani:2004,de2010bound,boutabia2005solution}.

This work aims to find the solutions of the DKP equation for spin-one particles under the hyperbolic tangential potential. Then, we calculate the reflection $R$ and transmission $T$ coefficients to analyze their behavior in the different regions of the potential. 

This work is organized in the following way. In Sec. \ref{sec_potential} we present the hyperbolic tangent potential, exploring its limiting case when $b\rightarrow\infty$. Section \ref{sec_DKP} is devoted to solving the DKP equation for particles of spin-one in presence of the hyperbolic tangent potential. Remarkably, the equation is reduced to a Klein--Gordon type equation. Sec. \ref{sec_results} shows the reflection and transmission coefficients, and the superradiance phenomena are discussed. Finally, in section \ref{sec_conclusion}, we present the conclusions of this work.

%%%%%%%%%%%%%%%%
\section{The hyperbolic tangent potential}
\label{sec_potential}
%%%%%%%%%%%%%%%%

\begin{figure}[htbp]
\begin{center}
\includegraphics[scale=0.50]{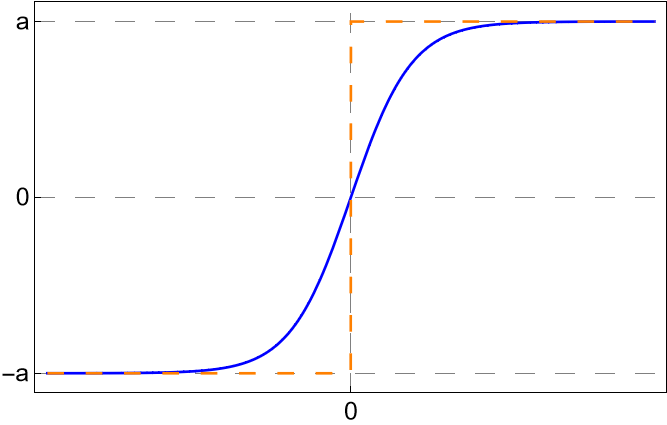}
\caption{Hyperbolic tangent potential for $a=1$ with $b=1$ (solid line), and $b=10000$ (dashed line)}
\label{fig_tanh}
\end{center}
\end{figure}

The hyperbolic tangent potential is a kind of smooth  barrier, and it is  defined by \cite{rojas:2015},

\begin{equation}
V(x)=a\,\tanh(b\,x),
\label{potential}
\end{equation}
where $a$ represents the height of the potential and $b$ gives the smoothness of the curve. Note that when $b \rightarrow \infty$ the hyperbolic tangent potential reduces to a step potential. This potential is represented in Fig. (\ref{fig_tanh}) for two different values of $b$. 

%%%%%%%%%%%%%%%%
\section{Scattering States}
\label{sec_DKP}
%%%%%%%%%%%%%%%

The DKP equation with the introduction of interaction with electromagnetic field\cite{chetouani:2004} is given by 

\begin{equation}
\label{DKP_spin}
\left[i \beta^\mu \left(\partial_\mu+i e A_\mu \right) -m\right]\Psi(x,t)=0,
\end{equation}

$\beta^\mu$ with $\mu= 0, 1, 2, 3$, are known as the DKP matrices and they satisfy the commutation relation\cite{hamzavi2013approximate,hassanabadi:2013a,hassanabadi:2013b}
\begin{equation*}
    \beta^\mu\beta^\nu\beta^\lambda+\beta^\lambda\beta^\nu\beta^\mu=g^{\mu\nu}\beta^\lambda+g^{\nu\lambda}\beta^\mu.
\end{equation*}
and the metric tensor is given by diag (1, -1, -1, -1). These matrices generate, as mentioned above, the DKP algebra. This algebra has three irreducible representations\cite{de2010bound}, a trivial one-dimensional representation, a five-dimensional representation, and a ten-dimensional representation. From which the second and third representations describe particles with spin--zero and spin--one, respectively. 

In particular, the beta matrices for spin-one are $10 \times 10$ dimension and are given by:

\begin{equation}
    \beta^{0}=
    \begin{pmatrix} 
	0  & \bar{0} & \bar{0} & \bar{0} \\
	\bar{0}^{T} &\textbf{0}&\textbf{1}&\textbf{0}\\
	\bar{0}^{T} &\textbf{1}&\textbf{0}&\textbf{0}  \\
	\bar{0}^{T} &\textbf{0}&\textbf{0}&\textbf{0} \\
	\end{pmatrix},
	\quad
	\beta^{i}=
    \begin{pmatrix} 
	0  & \bar{0} & e_i & \bar{0} \\
	\bar{0}^{T} &\textbf{0}&\textbf{0}&-is_i\\
	-e_{i}^{T} &\textbf{0}&\textbf{0}&\textbf{0}  \\
	\bar{0}^{T} &-is_i&\textbf{0}&\textbf{0} \\
	\end{pmatrix},
	\quad
    i= 1, 2, 3,
\end{equation}
where $s_i$ are $3 \times 3$ matrices which depending on an even permutation, an odd permutation, and repeated indices, it takes the form $(s_i)_{jk}=-i\epsilon_{ijk} $ respectively\cite{de2010bound}. Additionally, $\textbf{1}$ and $\textbf{0}$ are the $3 \times 3$ unity and zero matrices, $\bar{0}= (0,0,0)$, and $e_i$ are
\begin{equation*}
    e_1=(1,0,0), \hspace{3mm} e_2=(0,1,0), \hspace{3mm} e_3=(0,0,1).
\end{equation*}

Notice that our potential depends only on the $x$ component. Even more, as our potential does not depends on time, the solution of Eq. (\ref{DKP_spin}) can be written as $\Psi(x,t)=e^{iEt}\phi(x)$. Therefore, we can rewrite the DKP equation as

\begin{equation}\label{eq4}
    \left\{\beta^{0}\left[E-V(x)\right]+ i\beta^{1}\frac{d}{dx}-m \right\}\phi(x)=0,
\end{equation}
where $\phi(x)$ is a spinor which for spin-one particles will have $10$ components. So we might write it as: 
\begin{equation*}
    \phi(x)^{T}=(\psi_1, \psi_2, \psi_3, \psi_4, \psi_5, \psi_6, \psi_7, \psi_8, \psi_9, \psi_{10}).
\end{equation*}

In principle there are two possible paths to solve Eq. (\ref{eq4}), The first one is by introducing the full form of the $\beta^0$ and $\beta^1$ matrices, which will give us a system of 10 coupled equations. Nevertheless, by doing it, we will notice how we could group them. Hence, the second approach is by partitioning the DKP spinor into\cite{de2010bound}:
\begin{equation}
\begin{split}
    \Psi^{+}(x)=
        \begin{pmatrix}
           \varphi_{3} \\
           \varphi_{4} \\
         \end{pmatrix},
         \quad
    \Psi^{-}(x)=\varphi_{5},\\
    \Phi^{+}(x)=        
        \begin{pmatrix}
           \varphi_{6} \\
           \varphi_{7} \\
         \end{pmatrix},
         \quad
    \Phi^{-}(x)=\varphi_{2},\\
    \Theta^{+}(x)=
            \begin{pmatrix}
           \varphi_{10} \\
           \varphi_{-9} \\
         \end{pmatrix},
         \quad
    \Theta^{-}(x)=\varphi_{1}.
\end{split}
\end{equation}

And $\varphi_8=0$,Notice that this partition arrives from the 10 system of equations obtained if we solve the DKP equation by the first method. By using this partition, it is easy to show that  the one-dimensional time-independent DKP equation is reduced to the following system of equations.

\begin{equation}\label{eq6}
    \begin{cases}
        \left\{\dfrac{d^2}{dx^2}+[E-V(x)]^2-m^2\right\}\Psi^{\pm}=0,\\
        \Phi^{\pm}=\dfrac{1}{m}[E-V(x)]\Psi^{\pm},\vspace{1mm}\\
        \Theta^{\pm}=\dfrac{i}{m} \dfrac{d}{dx}\Psi^{\pm}.
    \end{cases}
\end{equation}

From Eq. (\ref{eq6}) it is easy to see that the second and third component depends on the first one\cite{chetouani:2004} and that the first component follows the Klein--Gordon type equation. 

Now, by replacing $V(x)=a\tanh(bx)$ we get

\begin{equation}\label{eq7}
    \left\{\dfrac{d^2}{dx^2}+\left[E-a\tanh(bx)\right]^2-m^2\right\}\Psi=0.
\end{equation}

By using the following change of variables $y=-e^{2bx}$ we reduce\cite{rojas:2015} Eq (\ref{eq7}) to

\begin{equation}\label{eq8}
    4b^2y\frac{d}{dy}\left(y\frac{d\Psi}{dy}\right)+\left\{\left[E+a\left(\frac{1+y}{1-y}\right)\right]^2-m^2\right\}\Psi=0.
\end{equation}

Now, we propose the following substitution\cite{rojas:2015} $\Psi=y^\alpha(1-y)^\beta f(y)$. It is not difficult to verify that we arrive to

\begin{equation}\label{eq9}
    y(1-y)f''(y)+[(2\alpha+1)-(2\alpha+2\beta+1)y]f'(y)+(\alpha+\beta-\gamma)(\alpha+\beta+\gamma)f(y)=0.
\end{equation}

Where the parameters $\alpha$, $\beta$ and $\gamma$ are define as

\begin{equation*}
    \begin{split}
        \alpha=i\nu, \hspace{2mm} \mathrm{and} \hspace{2mm} \nu =\frac{\sqrt{(E+a)^2-m^2}}{2b},\\
        \beta=\lambda, \hspace{2mm} \mathrm{and} \hspace{2mm} \lambda=\frac{b+\sqrt{b^2-4a^2}}{2b},\\
        \gamma=i\mu, \hspace{2mm} \mathrm{and} \hspace{2mm}\mu=\frac{\sqrt{(E-a)^2-m^2}}{2b}.
    \end{split}
\end{equation*}

Notice that Eq. (\ref{eq9}) has the form of a hypergeometric differential equation, which solution is known. 

\begin{equation}
\begin{split}
    \Psi(y) & = c1\hspace{1mm}y^\alpha(1-y)^\beta F_1(\alpha+\beta-\gamma,\alpha+\beta+\gamma,1+2\alpha,y)\\
    & + c2\hspace{1mm}y^{-\alpha}(1-y)^\beta F_1(-\alpha+\beta+\gamma,-\alpha+\beta-\gamma,1-2\alpha,y)\textbf{V},
\end{split}
\end{equation}
where \textbf{V} is a $3 \times 1$ vector needed to recover the three spin-one directions\cite{boutabia2005solution}. The next step is to plug this solution into the two other components of Eq. (\ref{eq6}). For that let us use the following hypergeometric function property
\begin{equation*}
    \frac{d}{dy}F_1(\alpha,\beta,\gamma,y)=\frac{\alpha\beta}{\gamma}F_1(\alpha+1,\beta+1,\gamma+1,y).
\end{equation*}

Therefore, we arrive at the following solution 

\begin{equation}\label{eq11}
    \begin{split}
        \begin{pmatrix}
           \Psi \\
           \Phi \\
           \Theta\\
         \end{pmatrix} =
         c1\hspace{1mm}y^\alpha(1-y)^\beta [F_1(\alpha_1,\beta_1,\gamma_1,y) M_1(y)+F_1(\alpha_1+1,\beta_1+1,\gamma_1+1,y)N_1(y)]\\
          c2\hspace{1mm}y^{-\alpha}(1-y)^\beta [F_1(\alpha_2,\beta_2,\gamma_2,y) M_2(y)+F_1(\alpha_2+1,\beta_2+1,\gamma_2+1,y)N_2(y)],
    \end{split}
\end{equation}
where; $\alpha_1=\alpha+\beta-\gamma$; $\beta_1=\alpha+\beta+\gamma$; $\gamma_1=1+2\alpha$ and; $\alpha_2=-\alpha+\beta+\gamma$; $\beta_2=-\alpha+\beta-\gamma$; $\gamma_2=1-2\alpha$. Even more, $M_1(y)$, $M_2(y)$, $N_1(y)$, $N_2(y)$ are $9 \times 1$ vectors given by;

\begin{equation}\label{eq12}
    \begin{split}
        M_1(y)=
        \begin{pmatrix}
           1 \\
          \dfrac{E}{m}-\dfrac{a(1+y)}{m(1-y)} \\
           \dfrac{2bi}{m}(1-y)^{-1}[\alpha-(\alpha+\beta)y]\\
         \end{pmatrix} \otimes \textbf{V}; \hspace{5mm}
         N_1(y)=
        \begin{pmatrix}
           0 \\
           0 \\
           \dfrac{2bi}{m}\dfrac{\alpha_1\beta_1}{\gamma_1}y\\
         \end{pmatrix} \otimes \textbf{V},\\
        M_2(y)=
        \begin{pmatrix}
           1 \\
          \dfrac{E}{m}-\dfrac{a(1+y)}{m(1-y)} \\
           \dfrac{-2bi}{m}(1-y)^{-1}[\alpha-(\alpha+\beta)y])\\
         \end{pmatrix} \otimes \textbf{V}; \hspace{5mm}
         N_2(y)=
        \begin{pmatrix}
           0 \\
           0 \\
           \dfrac{2bi}{m}\dfrac{\alpha_2\beta_2}{\gamma_2}y\\
         \end{pmatrix} \otimes \textbf{V}.\\
    \end{split}
\end{equation}

Notice that we arrive at two solutions, one is the incident wave and the other one is the reflected wave. Now, we need to construct the transmitted wave. For that let's use the following property. 

\begin{equation}\label{eq13}
\begin{split}
    F_1(a,b,c,z)=\dfrac{\Gamma(c)\Gamma(b-a)}{\Gamma(b)\Gamma(c-a)}(-z)^{-a}F_1(a,1-c+a,1-b+a,z^{-1})\\
    +\dfrac{\Gamma(c)\Gamma(a-b)}{\Gamma(a)\Gamma(c-b)}(-z)^{-b}F_1(b,1-c+b,1-a+b,z^{-1}).
\end{split}
\end{equation}
Hence, let's plug Eq. (\ref{eq13}) into our main solution Eq. (\ref{eq11}) to construct the transmitted wave. Therefore, we arrive to
%the transmitted wave becomes;

\begin{equation}\label{eq14}
    \begin{split}
        \begin{pmatrix}
           \Psi \\
           \Phi \\
           \Theta\\
         \end{pmatrix}_{\trans} =&
         c1(-1)^{i\nu}(-e^{2bx})^{-\lambda}(1+e^{2bz})^{\lambda}(e^{2bx})^{i\mu}\\&
         \big\{\left[\Gamma_1 F_1(\alpha_1,1-\gamma_1+\alpha_1,1-\beta_1+\alpha_1,-e^{-2bx})\right]M_1(x)\\&
        \\ &+\left[\Gamma_3(e^{-2bx})F_1(\alpha_1
+1,1-\gamma_1+\alpha_1,1-\beta_1+\alpha_1,-e^{-2bx})\right]N_1(x)\big\}\\&
          \\
         &+c2(-1)^{-i\nu}(-e^{2bx})^{-\lambda}(1+e^{2bz})^{\lambda}(e^{2bx})^{i\mu}\\&
         \\
         &\big\{\left[\Gamma_6 F_1(\beta_2,1-\gamma_2+\beta_2,1-\alpha_2+\beta_2,-e^{-2bx})\right]M_2(x)\\&  
         \\   &+\left[\Gamma_8(e^{-2bx})F_1(\beta_2+1,1-\gamma_2+\beta_2,1-\alpha_2+\beta_2,-e^{-2bx})\right]N_2(x)\big\}.
    \end{split}
\end{equation}

Now, as it is known the incident wave is equal to the sum of the transmitted and reflected waves. %%%% %%%%CITAR PAPER TANH KLEIN GORDON%%%%%%%%.
Then, by using once more the property (\ref{eq13}) on the transmitted wave we get

\begin{equation}\label{eq15}
\begin{split}
            \begin{pmatrix}
           \Psi \\
           \Phi \\
           \Theta\\
         \end{pmatrix}_{inc} =& (1+e^{2bx})^{\lambda}(e^{2bx})^{i\nu}
         \big[A F_1(\alpha_1,\beta_1,\gamma_1,-e^{2bx})M_1(x)\\&
         +B F_1(\alpha_1+1,\beta_1+1,\gamma_1+1,-e^{2bx})N_1(x)\big],
\end{split}
\end{equation}
\begin{equation}\label{eq16}
\begin{split}
         \begin{pmatrix}
           \Psi \\
           \Phi \\
           \Theta\\
         \end{pmatrix}_{\reff} = &(1+e^{2bx})^{\lambda}(e^{2bx})^{-i\nu}
         \big[C F_1(\alpha_2,\beta_2,\gamma_2,-e^{2bx})M_2(x)\\&
         D F_1(\alpha_2+1,\beta_2+1,\gamma_2+1,-e^{2bx})N_2(x)\big],
\end{split}
\end{equation}
where,

\begin{equation}\label{eq17}
    \begin{split}
        A=\dfrac{\Gamma(1-\beta_1+\alpha_1)\Gamma(1-\gamma_1)}{\Gamma(1-\gamma_1+\alpha_1)\Gamma(1-\beta_1)}, \hspace{5mm} B=\dfrac{\Gamma(1-\beta_1+\alpha_1)\Gamma(-\gamma_1)}{\Gamma(1-\gamma_1+\alpha_1)\Gamma(-\beta_1)},\\
        C=\dfrac{\Gamma(1-\alpha_2+\beta_2)\Gamma(1-\gamma_2)}{\Gamma(1-\gamma_2+\beta_2)\Gamma(1-\alpha_2)}, \hspace{5mm} D=\dfrac{\Gamma(1-\alpha_2+\beta_2)\Gamma(-\gamma_2)}{\Gamma(1-\gamma_2+\beta_2)\Gamma(-\alpha_2)}.
    \end{split}
\end{equation}\vspace{2mm}

To obtain the reflection and transmission coefficients, we need to study the asymptotic behavior of the solutions. In other words, let us explore the limiting case when $x\rightarrow\pm\infty$. 
Note that if $x\rightarrow\infty$ implies that $y\rightarrow-\infty$. On the other hand, when $x\rightarrow-\infty$ implies that $y\rightarrow0$.

First, for the incident wave, we have $(x\rightarrow-\infty)$

\begin{equation}\label{eq18}
    \begin{pmatrix}
           \Psi \\
           \Phi \\
           \Theta\\
         \end{pmatrix}_{inc} = A e^{2ib\nu x}
         \begin{pmatrix}
                1\\
                \dfrac{E-a}{m}\\\\
                \dfrac{2bi}{m}\alpha
         \end{pmatrix}\otimes \textbf{V}.
\end{equation}

By using,
\begin{equation*}
    \lim_{y\rightarrow0}(-y)^{i\nu}=e^{2ib\nu x}, \hspace{3mm} \lim_{y\rightarrow0}(1-y)^\lambda=1, \hspace{3mm} \lim_{y\rightarrow0}F_1(a,b,c,y)=1.
\end{equation*}\vspace{2mm}

Next, for the reflected wave $(x\rightarrow-\infty)$
\begin{equation}\label{eq19}
    \begin{pmatrix}
           \Psi \\
           \Phi \\
           \Theta\\
         \end{pmatrix}_{\reff} = C e^{-2ib\nu x}
         \begin{pmatrix}
                1\\
                \dfrac{E-a}{m}\\\\
                \dfrac{-2bi}{m}\alpha
         \end{pmatrix}\otimes \textbf{V}.
\end{equation}

By using,
\begin{equation*}
    \lim_{y\rightarrow0}(-y)^{-i\nu}=e^{-2ib\nu x}, \hspace{3mm} \lim_{y\rightarrow0}(1-y)^\lambda=1, \hspace{3mm} \lim_{y\rightarrow0}F_1(a,b,c,y)=1.
\end{equation*}\vspace{2mm}

Finally, for the transmitted wave $(x\rightarrow\infty)$

\begin{equation}\label{eq20}
    \begin{pmatrix}
           \Psi \\
           \Phi \\
           \Theta\\
         \end{pmatrix}_{\trans} = e^{2ib\mu x}
         \begin{pmatrix}
                1\\
                \dfrac{E+a}{m}\\
                \dfrac{2bi(\alpha+\beta)}{m}
         \end{pmatrix}\otimes \textbf{V}.
\end{equation}

By using,
\begin{equation*}
    \lim_{y\rightarrow-\infty}(-y)^{\lambda}=e^{-2ib\lambda x}, \hspace{3mm} \lim_{y\rightarrow-\infty}(1-y)^\lambda=e^{2b\lambda x}, 
\end{equation*}\vspace{-2mm}
\begin{equation*}
    \lim_{y\rightarrow-\infty}(-y)^{i\mu}=e^{2bi\mu x},\hspace{3mm}
    \lim_{y\rightarrow-\infty}F_1(a,b,c,y^{-1})=1.
\end{equation*}
The asymptotic behaviour of Eqs. (14), (15), and (16) are obtained studying the asymptotic behaviour of the hypergeometric functions in both limits $x \rightarrow \pm \infty$, which conduce to exponential functions that represents the incident, reflected and transmitted wave. This leads to the following conditions over the parameters $\alpha_1$, $\beta_1$, and $\gamma_1$

\begin{eqnarray}
\alpha_1&=&i\nu+\lambda-i\mu,\\
\beta_1&=&i\nu+\lambda+i\mu,\\
\gamma_1&=&1+2i\nu,
\end{eqnarray}
and over the parameters $\alpha_2$, $\beta_2$, and $\gamma_2$

\begin{eqnarray}
\alpha_2&=&-i\nu+\lambda+i\mu,\\
\beta_2&=&-i\nu+\lambda-i\mu,\\
\gamma_2&=&1-2i\nu.
\end{eqnarray}

From here, the process to obtain the reflection and transmission coefficients related to the hyperbolic tangential potential is straightforward.

Meanwhile, the components of the four-current $J^\mu$ can be calculated by\cite{de2010bound,cardoso2010nonminimal}

\begin{equation}
    J^1=\dfrac{1}{m}\Im \left[\Psi^{(+)\dagger} \dfrac{d\Psi^{(+)}}{dx}+\Psi^{(-)\dagger} \dfrac{d\Psi^{(-)}}{dx}\right],
\end{equation}
where $\Im$ means imaginary part of the function. Giving us as result the following currents.
 
\begin{align}
    J_{\inc}&=\dfrac{6AA^{*}b\nu}{m},\\
    J_{\reff}&=-\dfrac{6CC^{*}b\nu}{m},\\
    J_{\trans}&=\dfrac{6b\mu}{m}.
\end{align}

Then, it can be verified that the reflection and transmission coefficients are

\begin{align}\label{eq21}
        R&=\dfrac{|J_{\reff}|}{|J_{\inc}|}=\dfrac{\left|\dfrac{\Gamma(1-\alpha_2+\beta_2)\Gamma(1-\gamma_2)}{\Gamma(1-\gamma_2+\beta_2)\Gamma(1-\alpha_2)}\right|^2}{\left|\dfrac{\Gamma(1-\beta_1+\alpha_1)\Gamma(1-\gamma_1)}{\Gamma(1-\gamma_1+\alpha_1)\Gamma(1-\beta_1)}\right|^2},\\
        T&=\dfrac{|J_{\trans}|}{|J_{\inc}|}=\dfrac{\mu}{\nu}\frac{1}{\left|\dfrac{\Gamma(1-\beta_1+\alpha_1)\Gamma(1-\gamma_1)}{\Gamma(1-\gamma_1+\alpha_1)\Gamma(1-\beta_1)}\right|^2}.
\end{align}\vspace{2mm}

\begin{figure}[th!]
\begin{center}
\includegraphics[scale=0.50]{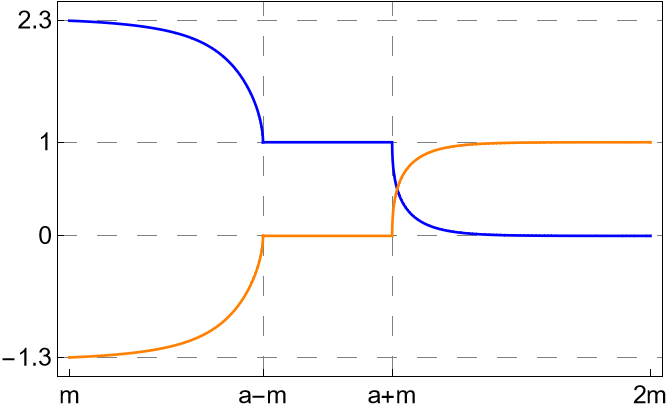}
\caption{The reflection $R$ and transmission $T$ coefficients varying energy $E$ for the hyperbolic tangent potential for $a=5$, $m=1$,  and $b=3$.}
\label{fig_RT}
\end{center}
\end{figure}

It is worth mentioning that the unitary relation $R+T=1$ is satisfied by the reflection and transmission coefficient obtained, as can be seen in Fig. \ref{fig_RT}. Both are given in terms of Gamma functions, and their behavior has been explored by using Wolfram Mathematica\textsuperscript{\textregistered} Software.

%%%%%%%%%%%%%%%%
\section{Superradiance}	
\label{sec_results}
%%%%%%%%%%%%%%%%

As the incident particle travels left to right, the relation dispersion $(\mu, \nu)$ has to be positive. Then, which defines the sign is the group velocity.\cite{calogeracos1999history}.  

\bigskip
\begin{align}
    \dfrac{dE}{d\nu}&=\dfrac{\nu}{E+a}\geq 0,\\
    \dfrac{dE}{d\mu}&=\dfrac{\mu}{E-a}\geq 0.
\end{align}
\bigskip

Therefore, according to the relation dispersion, the hyperbolic tangent potential can be split into five different regions.

As we can observe in Table 1, in the region where $a+m>E>a-m$ or $-a+m>E>-a-m$, the dispersion relation is imaginary. As in either of those regions, the transmitted wave does not have a real part (is attenuated); it implies that the wave is completely reflected $(R=1)$.

In addition, notice that in the region where $a-m>E>-a+m$, the dispersion relation behaves $\mu<0$ and $\nu>0$. This implies that our transmission coefficient $(T)$ has to be less than zero in that region. Consequently, to still satisfy the unitary condition, the reflection coefficient $(R)$ needs to be greater than one, producing the superradiance phenomenon in that region in the potential.

\begin{table}[th!]
\begin{center}
\begin{tabular}{|c|c|c|c|c|}
\hline
$E>a+m$ & $\nu>0$ & $\nu\in\mathbb{R}$ & $\mu>0$ & $\mu\in\mathbb{R}$ \\ \hline
$a+m>E>a-m$ & $\nu>0$ & $\nu\in\mathbb{R}$ & -- & $\mu\in\mathbb{I}$ \\ \hline
$a-m>E>-a+m$ & $\nu>0$ & $\nu\in\mathbb{R}$ & $\mu<0$ & $\mu\in\mathbb{R}$ \\ \hline
$-a+m>E>-a-m$ & -- & $\nu\in\mathbb{I}$ & $\mu<0$ & $\mu\in\mathbb{R}$ \\ \hline
$E<-a-m$ & $\nu<0$ &$\nu\in\mathbb{R}$ &  $\mu<0$  &  $\mu\in\mathbb{R}$ \\ \hline
\end{tabular}
\end{center}
\caption{The five regions of the hyperbolic tangent potential.}
\end{table}

\begin{figure}[th!]
\begin{center}
\includegraphics[scale=0.50]{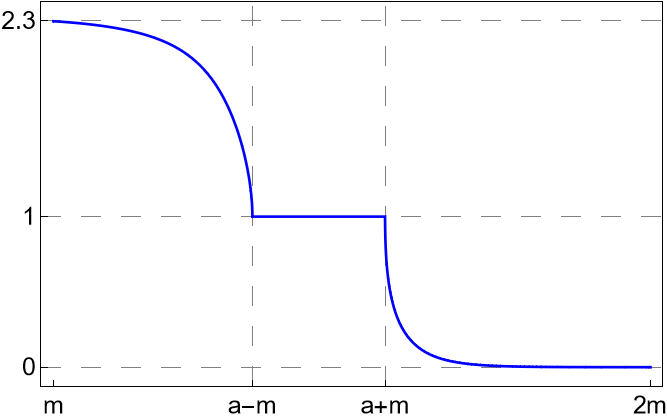}
\caption{The reflection coefficient $R$ varying energy $E$ for the hyperbolic tangent potential for $a=5$, $m=1$, and $b=3$.}
\label{fig_R}
\end{center}
\end{figure}

\begin{figure}[th!]
\begin{center}
\includegraphics[scale=0.50]{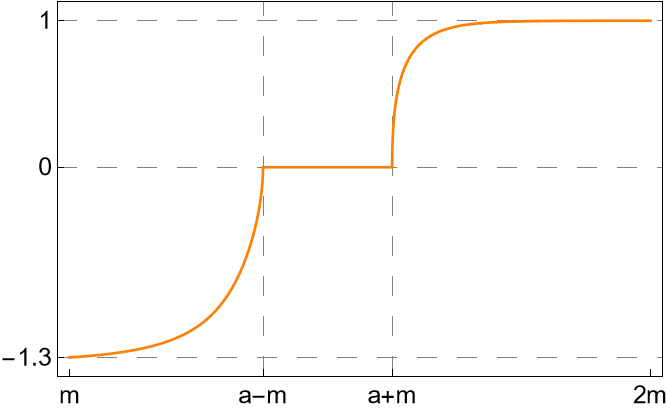}
\caption{The transmission coefficient $T$ varying energy $E$ for the hyperbolic tangent potential for $a=5$, $m=1$,  and $b=3$.}
\label{fig_T}
\end{center}
\end{figure}

Figures \ref{fig_R} and \ref{fig_T} show the reflection and transmission coefficients, respectively, for the hyperbolic tangent potential. We can observe that in the region $a-m>E>m$ is where the superradiance is produced.

\vspace{-2mm}

\newpage
%%%%%%%%%%%%%%%%
\section{Conclusions}	
\label{sec_conclusion}
%%%%%%%%%%%%%%%%

We solved the DKP equation for spin--one particles in presence of the hyperbolic tangent potential. By splitting and reorganizing the DKP spinor into three components, we show that the DKP equation is reduced to a Klein--Gordon equation with two more components, which depends on the first one. The reflections and transmission coefficients satisfy the unitary condition and are given in terms of Gamma functions. Even more, by using the dispersion relation, we demonstrated that for $a-m>E>m$ region, the transmission coefficient has to be less than zero and the reflection coefficient greater than one. Therefore, superradiance occurs in that region.

%%%%%%%%%%%%%%%%
%\bibliographystyle{unsrt}
%\bibliography{DKP.bib}

%%%%%%%%%%%%%%%%

\end{document}